\documentclass[10pt, conference,a4paper]{IEEEtran}
\usepackage[left=1.5cm,right=1.5cm,top=1.9cm, bottom=1.9cm]{geometry}
\usepackage{cite}
\usepackage{amsmath,amssymb,amsfonts}
\usepackage{algorithm2e}
\usepackage{balance}

\SetAlFnt{\small}
\SetAlCapFnt{\small}
\SetAlgoVlined
\RestyleAlgo{ruled}
\SetKwComment{Comment}{\# }{}
\SetKw{init}{\textbf{Initialization :}}
\usepackage{graphicx}
\usepackage{textcomp}
\usepackage{xcolor}
\usepackage{listings}
\usepackage{pdfpages}
\usepackage{bm}
\usepackage{cuted}%
\usepackage{footnote}
\usepackage{subcaption}
\usepackage[font=footnotesize]{caption}

\def\BibTeX{{\rm B\kern-.05em{\sc i\kern-.025em b}\kern-.08em
    T\kern-.1667em\lower.7ex\hbox{E}\kern-.125emX}}

\usepackage{circuitikz}
\usetikzlibrary{positioning}
\usepackage{pgfkeys}

\usepackage{pgfplots}

\pgfplotsset{compat=newest}
\pgfplotsset{plot coordinates/math parser=false}
\newlength\figureheight
\newlength\figurewidth
\usetikzlibrary[patterns]
\usetikzlibrary{shapes.symbols}

\usetikzlibrary{arrows.meta,positioning}
\tikzset{block/.style={draw, rectangle, fill=cyan!90,
        minimum height=2em, minimum width=3em},
    sum/.style={draw, circle, node distance=1cm},
    input/.style={coordinate},
    output/.style={coordinate},
    pinstyle/.style={pin edge={to-,thin,black}},
        saturation block/.style={%
            draw,
            path picture={
                \pgfpointdiff{\pgfpointanchor{path picture bounding box}{south west}}%
                {\pgfpointanchor{path picture bounding box}{north east}}
                \pgfgetlastxy\x\y
                \tikzset{x=\x*.4, y=\y*.4}
                \draw [very thin] (-1,0) -- (1,0) (0,-1) -- (0,1);
                \draw [very thick] (-1,-.7) -- (-.7,-.7) -- (.7,.7) -- (1,.7);
            },
        }
    }

\tikzset{%
        rateLimit block/.style={%
            draw,
            path picture={
                \pgfpointdiff{\pgfpointanchor{path picture bounding box}{south west}}%
                {\pgfpointanchor{path picture bounding box}{north east}}
                \pgfgetlastxy\x\y
                \tikzset{x=\x*.4, y=\y*.4}
                \draw [very thin] (-1,0) -- (1,0) (0,-1) -- (0,1);
                \draw [very thick] (-1,-1) -- (1, 1);
            },
        }
    }
    \usetikzlibrary{svg.path}
    \usepackage{scalerel}
    \usetikzlibrary{spy}

    \definecolor{orcidlogocol}{HTML}{A6CE39}
    \tikzset{
      orcidlogo/.pic={
        \fill[orcidlogocol] svg{M256,128c0,70.7-57.3,128-128,128C57.3,256,0,198.7,0,128C0,57.3,57.3,0,128,0C198.7,0,256,57.3,256,128z};
        \fill[white] svg{M86.3,186.2H70.9V79.1h15.4v48.4V186.2z}
                     svg{M108.9,79.1h41.6c39.6,0,57,28.3,57,53.6c0,27.5-21.5,53.6-56.8,53.6h-41.8V79.1z M124.3,172.4h24.5c34.9,0,42.9-26.5,42.9-39.7c0-21.5-13.7-39.7-43.7-39.7h-23.7V172.4z}
                     svg{M88.7,56.8c0,5.5-4.5,10.1-10.1,10.1c-5.6,0-10.1-4.6-10.1-10.1c0-5.6,4.5-10.1,10.1-10.1C84.2,46.7,88.7,51.3,88.7,56.8z};
      }
    }

    \newcommand\orcidicon[1]{\href{https://orcid.org/#1}{\mbox{\scalerel*{
    \begin{tikzpicture}[yscale=-1,transform shape]
    \pic{orcidlogo};
    \end{tikzpicture}
    }{|}}}}

    \usepackage[hidelinks]{hyperref}

\begin{document}





\title{Enhanced Low-Complexity Receiver Design \\for Short Block Transmission Systems}

\author{\IEEEauthorblockN{ Mody~Sy ${\orcidicon{0000-0003-2841-2181}}$\IEEEmembership{}~and~ Raymond~Knopp ${\orcidicon{0000-0002-6133-5651}}$}

       \IEEEauthorblockA{
        \text{EURECOM}, 06410 BIOT, France \\
        \textcolor{darkgray}{mody.sy@eurecom.fr,~raymond.knopp@eurecom.fr}}
}

%
\maketitle

\begin{abstract}
  This paper presents a comprehensive analysis and the  performance enhancement of short block length channel detection incorporating  training information. The current communication systems' short block length channel detection are assumed to typically consist of least squares channel estimation, followed by quasi-coherent detection. By investigating the receiver structure,  specifically the estimator-correlator, we show that the non-coherent term, which is often disregarded in conventional detection metrics, results in significant losses in terms of performance and sensitivity in typical operating regimes of 5G/6G systems. A comparison with the fully non-coherent receiver in multi-antenna configurations reveals substantial losses in low spectral efficiency operating areas. Additionally, we demonstrate that by employing an adaptive DMRS/data power adjustment, it is possible to reduce the performance loss gap  which is amenable to a more sensitive quasi-coherent receiver. However, both of the  aforementioned ML detection strategies can result in substantial computational complexity when processing long bit length codes.  We propose an  approach to tackle this challenge by introducing the principle of block/segment coding using First-Order RM Codes which  is amenable to low-cost decoding through block-based fast Hadamard transforms. The Block-based FHT has demonstrated to be cost-efficient with regards to decoding time, as it evolves from quadric to quasi-linear complexity with a manageable decline in performance. Additionally, by incorporating an adaptive DMRS/data power adjustment technique, we are able to bridge/reduce the performance gap with respect to the conventional maximum likelihood receiver and attain high sensitivity, leading to a good trade-off between performance and complexity to efficiently handle small payloads.
\end{abstract}

\section{Introduction}
The New Radio (NR) waveforms, such as {\em Physical Uplink Control Channel (PUCCH)}, have been specifically designed to transmit small payloads with minimal error rates in challenging signal-to-noise conditions.
To fully leverage their potential for {\em Ultra-Reliable Low Latency Communications} (URLLC) and their suitability for massive machine-type communications, it is crucial to develop robust coding strategies that enable low-complexity detection/decoding algorithms.\\
Therefore, there is a need for enhanced receiver designs that can accurately and reliably detect short data transmissions with low computational complexity and power consumption \cite{Lee2018}. The area of short block transmission has been thoroughly examined in the existing literature, covering various aspects such as the design of signal codes and the derivation of state-of-the-art converse and achievability bounds for both coherent and non-coherent communication \cite{Xhemrishi2019, Polyanskiy2010, Durisi2016, Ostman2019jrnal}.\\ This study aims to explore the enhancement of short-block length detection/decoding strategies. The main focus of this work is the baseline 3GPP PUCCH transmission, specifically the use of Reed-Muller codes paired with {\em Orthogonal Demodulation Reference Signals} (DMRS). The symbols, being message-independent, are used by the receiver to resolve channel uncertainty through explicit channel estimation or more advanced joint estimation and detection techniques.
Furthermore, through the structure of the receiver, namely the \emph{correlator-estimator}, we will show that the non-coherent term which is not typically used in conventional receivers can lead to a performance penalty which can have a significant effect on receiver sensitivity in the operating regimes of 5G/6G system. Additionally, in current communications systems, short block lengths detection/decoding  usually consists of least-squares channel estimation followed by quasi- coherent detection, but it should be noted that in both cases, the underlying algorithms are sub-optimal because of the detection procedure involving separate channel estimation and \emph{quasi-coherent detection}. Moreover, a \emph{fully non-coherent receiver} also incurs a substantial complexity cost, particularly when dealing with longer bit length transmissions, despite providing enhanced or acceptable detection performance in comparison to a \emph{conventional receiver}. Therefore, it becomes realistic to consider alternative detection/decoding strategies that offer a favorable performance/complexity trade-off.\\
 It is noteworthy that the use of {\em maximum likelihood} (ML) decoding for the above-mentioned receivers seems computationally demanding. Therefore, we propose an approach to tackle this challenge by introducing the principle of block/segment encoding using First-Order RM Codes which  is amenable to low-cost decoding through block-based {\em Fast Hadamard Transforms} (FHT). The Block-based FHT has demonstrated to be cost-efficient with regards to decoding time, as it evolves from quadric to quasi-linear complexity with a manageable decline in performance. Additionally, by incorporating an adaptive DMRS/data power adjustment technique, it is possible to bridge/reduce the performance gap  with respect to the conventional maximum likelihood receiver and attain high sensitivity, leading to a good trade-off between performance and complexity to efficiently handle small payloads.\\
The manuscript is structured as follows: Section II lays out the system model, the methodology is drawn up in Section III, Section IV presents the results and performance analysis, and finally Section V concludes the paper.
\section{System Model}
\begin{figure}[ht]
  \centering
  \input{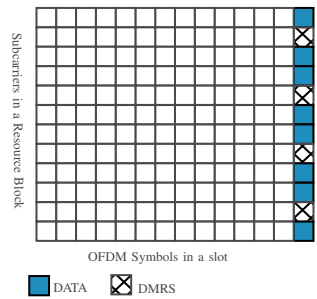}
  \caption{General resource mapping.}
  \label{fig:re_mapp_pucch2_1symb1}
\end{figure}

Consider a discrete-time model in which the transmitted and received symbols are $N$-dimensional column vectors, and thus a system is designed in such a way that the relationship between the transmitted and received signals   at the $i-th$ antenna port is as follows:
\begin{equation} \label{eqn:systmodel2}
\centering
    \mathbf{y}_i = {\mathbf h_i}\mathbf{x}+ \mathbf{z}_i, \quad i=0,1,\ldots,\mathsf{N_R}-1,
\end{equation}
where $\mathbf{y}_i$  represents an observed vector in $N$ complex dimensions, $\mathbf{x}$  is an $N$-dimensional modulated vector transporting $K$ channel bits, so that the message $\mathcal M = 0,1 \ldots, 2^K-1$, $\mathbf z_r$ is additive white Gaussian noise whose  real and imaginary components are independent and have variance $\sigma^2$ in each dimension.
$\mathsf{N_R}$ represents the number of observations of the transmitted vector over a multi-antenna receiver.
The transmitted vector $\mathbf{x}$ typically consists of data-independent components known as pilot or reference signals, so that $\mathbf{x} =\mathbf{x}^{(\mathsf d)}+ \mathbf{x}^{(\mathsf p)}$.
The subscripts $(\mathsf d)$ and $(\mathsf p)$ serve to denote the data components  and reference signals respectively.
 These reference signals serve to mitigate channel ambiguity in time, frequency, and space and are used to estimate the channel.
 In practice, the reference signals are {\em interleaved} among the data-dependent components to account for the characteristics of the propagation channel. This interleaving technique is commonly used in current OFDM systems.
 The number of data dimensions is denoted by $N_d$, and the number of reference signal dimensions is denoted by $N_p$, where $N_d+N_p=N$.
 In 3GPP, the standard notation for $N$ is $12\mathcal P \mathcal L$, where $\mathcal P$ refers to the number of {\em physical resource blocks} or (PRBs), each consisting of $12$ complex dimensions or resource elements. The value of $\mathcal P$ usually falls within a range of $1$ to $16$. $\mathcal L$ represents the number of symbols, typically ranging from $1$ to $14$, but it can be increased if multiple slots are utilized to signal the $K$ bits.
To demonstrate the consequences of disregarding the non-coherent term, PUCCH format 2 will be employed as an illustration. This particular format, based on OFDM, is a short PUCCH format and has the capability to transmit more than two bits utilizing either one or two OFDM symbols\cite{3GPP38211}. The transmission framework exhibits a simple procedure. Prior to transmitting short block channels, ranging in length from 3 to 11 bits, the messages are subjected to encoding utilizing a (32, $K$) Reed-Muller coding scheme.
$c_{\ell}=\displaystyle\left(\sum_{k=0}^{K-1} b_{k} \cdot M_{\ell, k}\right)\operatorname{mod} 2$,
where $\ \ell=0, \ 1, \ \ldots, \ N^\prime-1$ , $N^\prime $= $32$ and $M_{\ell, k}$ represents the basis sequences as defined in \cite{3GPP38212}.
It turns out that a classical $(N^\prime, K)$ block code is utilized among the family of Reed-Muller codes with a block length of 32, denoted as $N^{\prime}$. The information bitwidth is defined as $K$. The input bit sequence to the encoder is $b(0), b(1), b(2), \ldots, b(K-1)$ and the resulting sequence before rate matching is represented as $c(0), c(1), c(2), \ldots, c(N^\prime-1)$. The output bit sequence, following rate matching, is denoted as $e(0), \ e(1), \ e(2), \ \ldots, \ e(E-1)$, where the length of the rate matching output sequence, $E$, is dependent on the number of PRBs. The rate matching process in this case is simply repetition. The encoded {\em Uplink Control Information} (UCI) payload, denoted as $\mathbf e$, undergoes a scrambling process prior to modulation, resulting in a new block of bits $\tilde{\mathbf e}$ such that $\tilde{e}(\ell)=[e(\ell)+d(\ell)] \operatorname{mod} 2$.
The scrambling sequence $d(\ell)$ is a pseudo-random sequence derived from {\em radio resource control} (RRC) configuration parameters. The $\tilde{\mathbf e}$ bits are then subjected to  {\em quadrature phase shift keying} (QPSK) modulation, resulting in a block of complex-valued modulation symbols $x(0), \ x(1), \ \ldots, \ x\left(E / 2-1\right)$. The resource mapping process follows, whose aim is to allocate the modulated symbols onto resource occasions in both time and frequency domains and inserting DMRS resources. As illustrated in Figure~\ref{fig:re_mapp_pucch2_1symb1}, the resource mapping  here is embedded in the same spirit as in 3GPP PUCCH2 transmission.
On reception, the messages are recovered using maximum likelihood decoding, which can be computationally demanding.
\section{Methodology}
\subsection{Maximum Likelihood Receivers}
We describe detection metric   for  classical non-coherent channel  with unknown phase. The  LOS channel is $\mathbf{h}_i=\exp{\left(j\theta_i\right)}\mathbf{I}$ where $\theta_i$ are assumed to be i.i.d. uniform random variables on $[0,2\pi)$ .
We denote the likelihood function for the observed vector on a particular receiver branch with respect to a given transmitted signal as
\begin{equation}\label{eqn:lfpcsi}
\begin{aligned}
q\left(\mathbf{x},\left\{\mathbf{y}_i\right\}\right)=p\left(\left\{\mathbf{y}_i\right\} \mid \mathbf{x}\right)
=p\left(\left\{\mathbf{y}_i\right\} \mid \mathbf{x}, {\theta}_i\right)p\left({\theta}_i\right),
\end{aligned}
\end{equation}

where we have the fact que $p(\theta_i)= \frac{1}{2\pi}$,  is a uniform distribution on $[0, 2\pi)$.
The likelihood function is equivalent to
\begin{equation}
\begin{gathered}
q\left(\mathbf{x},\left\{\mathbf{y}\right\}\right)=\prod_{i=0}^{\mathsf{N_R}-1}\frac{1}{2\pi}\displaystyle \int_{{\theta_i}=0}^{2\pi} \exp \left(-\frac{\lvert|\mathbf y_{i}-\mathbf h_i \mathbf x\rvert|^2}{N_0}\right)\mathrm{d} \theta_i .
\end{gathered}
\end{equation}
Moreover we have
\begin{align}
\lvert |\mathbf{y}_{i} -  \mathbf h_i  \mathbf x \rvert |^2 &=\lvert | \mathbf{y}_{i} \rvert |^2 +  \lvert |\mathbf h_i  \mathbf x \rvert |^2 - 2\mathrm{Re}\left(\mathbf y_{i}\mathbf h_i^* \mathbf x^*\right).
\end{align}
Considering that  $ \mathbf x^\dag \mathbf y_{i} = \left|\mathbf x^\dag \mathbf y_{i}\right|e^{j\phi_i}$ where $\phi_i =  \angle{\mathbf x^\dag \mathbf y_i}$, the likelihood function subsequent to the exclusion of multiplicative terms that are unrelated to $\mathbf x$ can be expressed as:
\begin{equation}
\begin{aligned}
&q\left(\mathbf{x},\left\{\mathbf{y}\right\}\right)\propto\prod_{i=0}^{\mathsf{N_R}-1}\exp \left( - \frac{\left\|\mathbf{x}\right\|^2}{N_0}\right)\cdot\\&\frac{1}{2\pi}\int_{0}^{2\pi}\exp \left(\frac{2 }{N_0}\left|\mathbf x^\dag \mathbf y_{i}\right|\cos{\left(\phi_i -\theta_i\right)}\right)\mathrm{d} \theta_i,
\end{aligned}
\end{equation}
where $\displaystyle \frac{1}{\pi}\int_{{\lambda}=0}^{\pi}\exp(zcos(\lambda))\mathrm{d} \lambda=\operatorname{I_0(z)}$,  $\operatorname{I_0(\cdot)}$ is the  modified  bessel function of the first kind and $^\dag$ denote the complex conjugate transpose or Hermitian.
\begin{equation}\label{eqn:llr}
\begin{aligned}
q\left(\mathbf{x},\left\{\mathbf{y}\right\}\right)
&=\prod_{i=0}^{\mathsf{N_R}-1}\exp \left( - \frac{\left\|\mathbf{x}\right\|^2}{N_0}\right)\operatorname{I_0}\left(\frac{2 }{N_0}\left|\mathbf{x}^{\dag}\mathbf{y}_i\right|\right).
\end{aligned}
\end{equation}
We typically simplify the Likelihood fonction from (\ref{eqn:llr}) via a \emph{max-log approximation}.
The approximation is obtained by assuming that $\log \left\{\sum_{i} \exp \left(\lambda_{i}\right)\right\} \sim \max {i}\left\{\lambda{i}\right\}$, taking into account the exponential approximation of the modified Bessel function of the first kind, $I_{0}(z) \sim \frac{e^{z}}{\sqrt{2 \pi z}} \sim e^{z}$.

Hence, the maximum likelihood detection metric of the input  signal  can be closely approximated as:
\begin{equation}\label{eqn:mlr}
\hat{\mathbf x}=\underset{\mathbf x}{\operatorname{argmax}}\left(\sum_{i=0}^{\mathsf{N_R}-1}  \frac{2}{N_0}\left|\mathbf{x}^{\dag}\mathbf{y}_i\right|-\frac{|| \mathbf{x}||^2}{N_0}\right).
\end{equation}
It is worth mentioning that, in (\ref{eqn:mlr}), many of the terms can be omitted  when $||\mathbf{x}||$ is constant, as would be the case for BPSK or QPSK modulation for instance.
\begin{equation}\label{eqn:mlr_simpl}
\hat{\mathbf x}=\underset{\mathbf x}{\operatorname{argmax}}\displaystyle\sum_{i=0}^{\mathsf{N_R}-1}  \left|\mathbf{x}^{\dag}\mathbf{y}_i\right|
=\underset{\mathbf x}{\operatorname{argmax}}\displaystyle \sum_{i=0}^{\mathsf{N_R}-1}  \left|\mathbf{x}^{\dag}\mathbf{y}_i\right|^2.
\end{equation}
The form of the metrics has been observed to be distinct, with no overlap between the data and DMRS symbols.
By writing $\mathbf x = \mathbf x^{(\mathsf p)} + \mathbf x^{(\mathsf d)}$, then %
\begin{equation}
\left|\left(\mathbf{x}^{(\mathsf d)^\dag}+\mathbf{x}^{(\mathsf p)^\dag}\right) \mathbf{y}_i\right|^2=\left|\mathbf{x}^{(\mathsf p)^{\dag}} \mathbf{y}_i^{(\mathsf p)}+\mathbf{x}^{(\mathsf d)^{\dag}} \mathbf{y}_i^{(\mathsf d)}\right|^2
\end{equation}
More generally, in the traditional literature,  the detection of signals with training information can be extended using (\ref{eqn:mlr_simpl}) which is often referred to as an estimator-correlation. The equation encompasses both a quasi-coherent detection term and a non-coherent energy term.
\begin{equation}\label{eqn2.5}
\begin{aligned}
 \hat{\mathbf{x}}^{(\mathsf d)}&=\underset{\mathbf{x}^{(\mathsf d)}}{\operatorname{argmax}} \underbrace{\sum_{i=0}^{\mathsf{N_R}-1}\left|\mathbf{x}^{(\mathsf p)^\dag} \mathbf{y}_i^{(\mathsf p)}\right|^2}_{\text {data-independent term }}+\underbrace{\sum_{i=0}^{\mathsf{N_R}-1}\left|\mathbf{x}^{(\mathsf d)^\dag} \mathbf{y}_i^{(\mathsf d)}\right|^2}_{\text {non-coherent data term }}+ \\
& \hspace{5.5em}\underbrace{2 \operatorname{Re}\left(\sum_{i=0}^{\mathsf{N_R}-1} \mathbf{x}^{(\mathsf p)^\dag} \mathbf{y}_i^{(\mathsf p)} \cdot  \mathbf{y}_i^{(\mathsf d)^\dag}\mathbf{x}^{(\mathsf d)}\right)}_{\text {quasi-coherent term }},
\end{aligned}
\end{equation}

where $  \mathbf{x}^{(\mathsf p)^\dag} \mathbf{y}_i^{(\mathsf p)}  = \left(\mathbf{x}^{(\mathsf p)^\dag} \mathbf{x}^{(\mathsf p)}\right)\hat{\mathbf h}_i = N_p \hat{\mathbf h}_i$.

We demonstrate that neglecting the non-coherent energy term in current systems leads to a significant performance penalty in typical operating regimes of 5G/6G systems. This is contrary to the widespread belief that the impact of non-coherent energy term is negligible.
\subsection{Low  Complexity Receiver Design}
Reed-Muller codes (RM codes) are commonly known to be decodable using Hadamard or Fast Hadamard transforms. However, it is well-established that decoding the first-order RM code ($RM(r=1, m)$) using FHT is easier compared to higher-order RM codes ($r\geq 2$). In recent literature, several innovative algorithms have been proposed for decoding RM codes of any order \cite{Dumer2006, Ye2019, Li2021, Lian2020, Doan2022}.\\
Although maximum likelihood decoding algorithms have been extensively investigated in traditional literature for decoding data packets encoded with first-order Reed-Muller codes \cite{Ashikhmin}, it can become computationally expensive when the message length exceeds 6 bits. This is because the resulting codewords tend to be excessively long, leading to complex decoding processes that involve high-dimensional Hadamard transforms when using a FHT-based decoder.
This presents a significant challenge for transmitting short packets as the cost can be substantial. As an illustration, for a message of $K=11$ bits, the length of the code words would be $N^\prime=2^{10}$  bits using a first order $RM(1, m=10)$. Hence, to address the constraint of having a message length of $K\geq 6$ bits, we can utilize the principle of encoding and decoding by blocks. This method takes advantage of the low complexity decoding offered by FHT-based decoders. The objective is to segment the message into smaller, more manageable segments of bits, which can then be fed into $RM(1,m)$ encoders and concatenated. Upon reception, the received code is deconcatenated and decoded through the appropriate dimension of the Hadamard transform which is amenable to a low complexity receiver.
\subsubsection{ Block-based Encoding Principle}~\\
In instances where the payload exceeds $6$ bits, such as in the case of $K = 11$ bits, a combination of two first-order RM codes, $RM(1, m=4)$ and $RM(1, m=5)$, can be employed to encode the respective sub-blocks of $5$ bits and $6$ bits.\\ In regards to the Reed-Muller code $RM(1, 4)$, the codewords are generated using (\ref{eqn:enc_sblk1})
\begin{equation}\label{eqn:enc_sblk1}
\mathbf{c}^{(1)}=\mathbf{m}^{(1)}{\mathbf G}^{(1)}= \mathbf m^{(1)} . \left[\begin{array}{c}
\mathbf{1}\quad\mathbf{v}_4\quad\mathbf{v}_3\quad\mathbf{v}_2\quad\mathbf{v}_1
\end{array}\right]^\mathrm T .
\end{equation}
This code is characterized as an $(N^\prime=16, K=5, d_{min}=8)$ code, where the minimum distance of the $RM(r,m)$ is defined as $2^{m-r}$. The monomials of degree less than or equal to $r$ are represented by ${\mathbf 1, \mathbf  v_1, \mathbf v_2, \mathbf v_3, \mathbf v_4}$, with associated vectors, as indicated in the table below.
\begin{table}[htbp]
  \centering
    \begin{tabular}{r|r|r|r|r}
    \hline
    {$\mathbf 1$} & \multicolumn{1}{l|}{{$\mathbf v_4$}} & \multicolumn{1}{l|}{{$\mathbf v_3$}} & \multicolumn{1}{l|}{{$\mathbf v_2$}} & \multicolumn{1}{l|}{{$\mathbf v_1$}} \\
    \hline
    \hline
    1     & 0     & 0     & 0     & 0 \\
    \hline
    1     & 0     & 0     & 0     & 1 \\
    \hline
    1     & 0     & 0     & 1     & 0 \\
    \hline
    1     & 0     & 0     & 1     & 1 \\
    \hline
    1     & 0     & 1     & 0     & 0 \\
    \hline
    1     & 0     & 1     & 0     & 1 \\
    \hline
    1     & 0     & 1     & 1     & 0 \\
    \hline
    1     & 0     & 1     & 1     & 1 \\
    \hline
    1     & 1     & 0     & 0     & 0 \\
    \hline
    1     & 1     & 0     & 0     & 1 \\
    \hline
    1     & 1     & 0     & 1     & 0 \\
    \hline
    1     & 1     & 0     & 1     & 1 \\
    \hline
    1     & 1     & 1     & 0     & 0 \\
    \hline
    1     & 1     & 1     & 0     & 1 \\
    \hline
    1     & 1     & 1     & 1     & 0 \\
    \hline
    1     & 1     & 1     & 1     & 1 \\
    \hline
    \end{tabular}%
  \label{tab:addlabel}%
\end{table}%

If we consider $RM(1, 5)$,  code words are generated by
\begin{equation}
\mathbf{c}^{(2)}=\mathbf{m}^{(2)}{\mathbf G}^{(2)}= \mathbf m^{(2)} . \left[\begin{array}{c}
\mathbf{1}\quad\mathbf{v}_5\quad\mathbf{v}_4\quad\mathbf{v}_3\quad\mathbf{v}_2\quad\mathbf{v}_1
\end{array}\right]^\mathrm T.
\end{equation}
This is a $(N^\prime=32, K=6, d_{min}=16)$ code.
Therefore, the process of concatenation entails the merging of the two sub-codes,
\begin{equation}
\mathbf c=[\mathbf c^{(1)} \ \mathbf c^{(2)}] = [\mathbf m^{(1)}\mathbf G^{(1)} \ \mathbf m^{(2)}\mathbf G^{(2)} ].
\end{equation}
\subsubsection{ Block-based Decoding via FHT}~\\
Consider the received sequence $\mathbf u = (u_0,u_1, \ldots, u_{2^m-1}) \in \mathbb{F}_2$, and let $\mathbf c = (c_0,c_1, \ldots, c_{2^m-1})\in \mathbb{F}_2$ be a codeword. The bipolar representation of $\mathbf u$ is denoted as $\mathbf U \in \{-1,+1\}$ and is defined as $\mathbf U= (-1)^{\mathbf u}$. $\mathbb F_2$ denotes the Galois field.
Similarly, the bipolar representation of $\mathbf c$ is denoted as
$\mathbf C$ and defined as $\mathbf C= (-1)^{\mathbf c}$.
The decoding algorithm involves computing the correlation between $\mathbf U$ and $\mathbf C_i$, denoted as $\Delta_i$, for each of the $2^{m-1}$ codewords $\mathbf C_i= (-1)^{\mathbf c_i}$. The final step is to select the codeword for which $\Delta_i$ is the maximum. The simultaneous computation of all correlations can be depicted as a matrix representation. Denoting the column vector $C_i$ and constructing the matrix
$\mathbf H=\left[\begin{array}{llll}\mathbf{C}_0 & \mathbf{C}_1 & \ldots & \mathbf{C}_{2^{m}-1}\end{array}\right]$, the computation of all correlations can be expressed as follows:
\begin{equation}
\mathbf{\Delta}=\mathbf{U} \mathbf H .
\end{equation}
Where $\mathbf H$ is a Hadamard matrix of dimension $2^m$.
For the first subblock, which utilizes a $RM(1, 4)$ code with generator matrix $\mathbf G^{(1)}$,  $\mathbf H^{(1)}_{16}$ is employed. Similarly, for the second subblock, which employs a $RM(1, 5)$ code with generator matrix $\mathbf G^{(2)}$, $\mathbf H^{(2)}_{32}$ is utilized.

\begin{figure*}[!ht]
        \centering
        \includegraphics[width=0.8\linewidth]{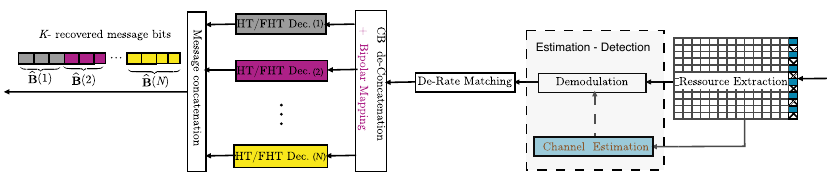}
        \caption{Block-based HT/FHT based-decoding of Short blocks : Receiver end. }
       \label{fig:rmfirt_order_rx}
\end{figure*}

 Furthermore, comprehensive descriptions of algorithms for first order RM code decoding using the Hadamard transform can be found in the works of Moon \cite{Moon2005} and Wicker \cite{Wicker94}.
This decoding process  can be optimized through the utilization of a FHT which is applicable to Hadamard matrices produced through the Sylvester construction. This optimization is based on the fact that $\mathbf H_{2^m}=\mathbf H_2 \otimes \mathbf H_{2^{m-1}}$, where the Kronecker product of matrices, denoted by $\otimes$, is applied. As a result, the matrix $\mathbf H_{2^m}$ can be decomposed as stated in the theorem derived from linear algebraic principles\cite{Moon2005}.
\begin{equation}
\begin{aligned}
& \mathbf H_{2^m}=\mathbf W_{2^m}^{(1)} \mathbf W_{2^m}^{(2)} \cdots \mathbf W_{2^m}^{(m)}.
\end{aligned}
\end{equation}
where $\mathbf W_{2^m}^{(i)}=\mathbf I_{2^{m-i}} \otimes \mathbf H_2 \otimes \mathbf I_{2^{i-1}}$, $\mathbf I$ is an identity  matrix.
\\
 Thus it comes,
\begin{equation}
\begin{aligned}
\mathbf H^{(1)}_{16}&=\mathbf W_{16}^{(1)} \mathbf W_{16}^{(2)} \mathbf W_{16}^{(3)} \mathbf W_{16}^{(4)}\\
&=\left(\mathbf I_{2^3} \otimes \mathbf H_2 \otimes \mathbf I_{2^0}\right)\left(\mathbf I_{2^2} \otimes \mathbf H_2 \otimes \mathbf I_{2^1}\right)\\&\quad\left(\mathbf I_{2^1} \otimes \mathbf H_2 \otimes \mathbf I_{2^2}\right)\left(\mathbf I_{2^0} \otimes \mathbf H_2 \otimes \mathbf I_{2^3}\right).
\end{aligned}
\end{equation}
Let's consider $\mathbf U^{(1)} = [U_0, U_1,\ldots , U_{15}]$, the first received sequence to be fed to the first decoder.
The corresponding Hadamard transforms can then be written as
\begin{equation}
 \mathbf{\Delta}^{(1)}=\mathbf{U}^{(1)} H^{(1)}_{16}=\mathbf{U}^{(1)}\left(\mathbf W_{16}^{(1)} \mathbf W_{16}^{(2)} \mathbf W_{16}^{(3)} \mathbf W_{16}^{(4)}\right),
\end{equation}
where
\begin{equation}
\begin{aligned}
\mathbf W_{16}^{(1)}&=\mathbf I_8 \otimes \mathbf H_2\\
\mathbf W_{16}^{(2)}&=\mathbf I_4 \otimes \mathbf H_2 \otimes \mathbf I_2\\
\mathbf W_{16}^{(3)}&=\mathbf I_2 \otimes \mathbf H_2 \otimes \mathbf I_4\\
\mathbf W_{16}^{(4)}&=\mathbf H_2 \otimes \mathbf I_8.
\end{aligned}\nonumber
\end{equation}
The conventional computation of the Hadamard transform $\mathbf H_{2^m}$ results in $2^m$ elements, each of which is obtained through $2^m$ addition/subtraction operations. This leads to a computational complexity of ${(2^m)}^2=\mathcal O\left({N^\prime}^2\right)$, which is equivalent to the complexity of a standard ML decoder that operates in a quadratic order. In contrast, the FHT, which has $m$ stages, has a computational complexity of $m2^m=\mathcal O\left(N^\prime\log N^\prime\right)$(i.e., {\em quasi-linear complexity}) due to its $2^m$ addition/subtraction operations per stage.
 The procedure of block-based RM decoding is therefore depicted in Figure \ref{fig:rmfirt_order_rx}.

\section{Numerical Results}
For illustration purposes, we focuse on PUCCH2-based short block lengths. PUCCH2 is configurable in terms of the resource usage, but we consider the simplest comprising of $2$ groups of 12 dimensions or {\em resource elements}, so-called  PRBs, making $24$ dimensions which consist of $16$ for data components, and $8$ for  DMRSs, which are known symbols used for channel estimation and tracking.
The simulations were performed under the assumption of TDL-C Non-Line-Of-Sight $300\ ns$ wireless channel, utilizing 2, 4 and 8 antenna configurations.  The TDL-C  channel model, which adheres to the 3GPP reference specifications, was employed. This model is distinguished by its extended time delay spread and its primary focus on non-MIMO assessments\cite{3GPP38901}. The antenna ports were subjected to independent and identically distributed realizations, with no incorporation of correlation modeling.

\begin{figure*} [!ht]
    \centering
  \subfloat[\scriptsize {
  Performance loss w.r.t Non-coherent term.}
  \label{Fig2_TDL_pucch2_Performance_loss}]{
   \includegraphics[width=0.46\linewidth]{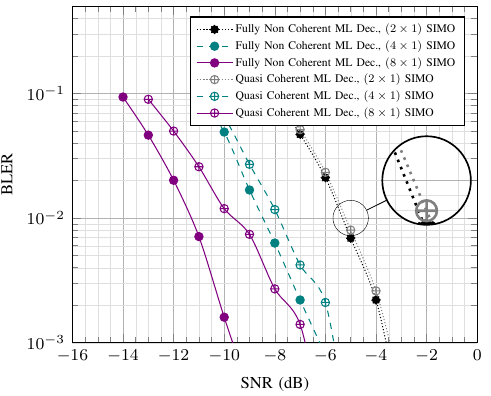}
  }
  \subfloat[\scriptsize {
  Adaptive DMRS power Adjustment via $\beta$.
}\label{fig:adaptative_power_ajustement_tdlc}]{%
    \includegraphics[width=0.46\linewidth]{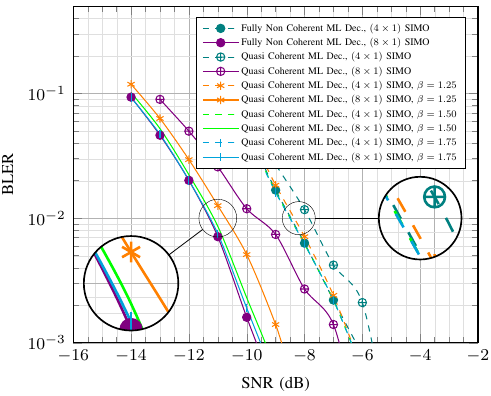}
    }
  \caption{Block Error Rate, 4 bits, ($\{8, 4, 2\} \times 1$) SIMO, 3GPP TDL-C urban NLOS channel, $300$ns of Delay spread, Unknown {\em Channel State Information}.}
  \label{fig:extra_enh}
\end{figure*}

In Figure~\ref{Fig2_TDL_pucch2_Performance_loss}, the comparison of performance between the optimal ML receiver, which is a {\em fully non-coherent detector}, and the conventional receiver, which is a quasi-coherent detector, is presented. The difference in performance arises due to the fact that the {\em non-coherent energy term} is disregarded in the quasi-coherent detector.




\begin{figure}[!ht]
        \centering
        \includegraphics[width=0.94\linewidth]{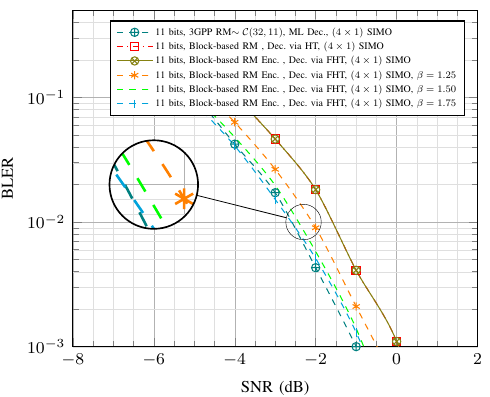}
        \caption{Block Error Rate, 11 bits, ($4 \times 1$) SIMO, 3GPP TDL-C urban NLOS channel, $300$ns of Delay spread, Unknown {\em Channel State Information}, Adaptive DMRS power Adjustment via $\beta$. }
        \label{fig:adaptative_power_ajustement11bits}
\end{figure}

In Figure~\ref{Fig2_TDL_pucch2_Performance_loss}, it is demonstrated that as the BLER threshold is set to $1\%$, the performance losses incurred by the penalties increase. This is evidenced by the estimated performance losses of $0.2$ dB, $0.6$ dB and $1.8$ dB for the $2$, $4$ and $8$ receive antenna configurations, respectively. Furthermore, it can be deduced that the performance gap expands with the increase in the number of receiving antennas.
This reflects that operating in regions of low spectral efficiency while disregarding the {\em non-coherent term} leads to a significant degradation in performance and a decrease in receiver sensitivity. These regions are defined by a low signal-to-noise ratio (SNR), which greatly affects the precision of channel estimates in each receiving chain.
Consequently, the significance of the \emph{non-coherent energy term} becomes increasingly evident. From this insight, the quasi-coherent receiver is sub-optimal in terms of its performance, despite its reduced complexity.\\ The observed decline in performance could potentially be remediated through the enhancement of channel estimate accuracy.\\
Furthermore, it is also conceivable to address the performance discrepancies observed in comparison to the {\em fully non-coherent receiver} by employing an adaptive DMRS/data power adjustment.
 In instances where we transmit reference and data symbols jointly in common OFDM symbols, as is the case for PUCCH or PUSCH or in some downlink control channels, there lies the potential to adjust the power ratio between data and reference signals.
This is applicable to both downlink and uplink transmissions without incurring any penalty in terms of peak power increase. Therefore, the system can be conceived as $\mathbf{y}_i = \left(\mathbf{x}^{(\mathsf d)} + \beta\cdot\mathbf{x}^{(\mathsf p)}\right)\mathbf{h}_i + \mathbf{z}_i$.
This adaptive power adjustment procedure is contingent on the values of $\beta$. The DMRS Power is to be slightly increased in a judicious manner since $\beta$ must be perfectly calibrated to ensure compliance with potential radio frequency constraints.
As depicted in Figure~\ref{fig:adaptative_power_ajustement_tdlc}, the performance improvement can be observed as a function of varying values of $\beta$.
Optimal performance improvement is achieved when $\beta$ is set to $1.75$. As demonstrated through simulation, when nRx=$4$ and $\beta$ is selected as $1.5$, the sub-optimal ML receiver exhibits comparable performance to the optimal ML receiver, effectively bridging the initial $0.6$ dB gap at the {\em Bit Error Rate} (BLER) threshold of $1\%$.
Furthermore, as we stated earlier, message recovery using maximum likelihood decoding is computationally demanding for both {\em fully non-coherent} and {\em quasi-coherent} receivers. Therefore, through Figure~\ref{fig:adaptative_power_ajustement11bits}, we provide some simulation results to show the merit of the proposed block-based decoding via  HT/FHT with respect to the recovery of information bits using a quasi-coherent ML decoder.
The results suggest that the ML decoder exhibits improved performance, however, this advantage is accompanied by a significant increase in computational complexity, even when utilizing the block decoding principle. 
In practical implementation, each received code block ($\mathcal{C}(11,48)=[ \mathcal{C}_1(5,16), \mathcal{C}_2(6,32)$)]) is processed through a corresponding  decoder, followed by concatenation at the decoder output stage. It is important to acknowledge that the efficacy of block decoding through both HT and FHT is comparable, exhibiting a marginal deviation of $1$ dB from the {\em Maximum Likelihood} (ML) receiver at a BLER threshold of $0.1\%$.
In terms of computational efficiency, block decoding via Hadamard transform demonstrates superior performance with a more expeditious convergence time. Consequently, it is advisable to employ this technique in scenarios where time sensitivity is a critical factor, specifically in URLLCs and sporadic random access applications.\\
Additionally, the performance disparity between the ML receiver and FHT based-receiver can be mitigated via the adaptive power adjustment procedure decribed above.
When the number of receive antennas is $4$ and $\beta$ is selected as $\beta = 1.75$ , the performance of the FHT-based decoder approaches that of the ML decoder at a BLER threshold of $1\%$, which corresponds to $1$ dB in terms of additional gain. This result is significant as it indicates that the FHT-based decoder provides a better trade-off between complexity and performance for short packet decoding, owing to its lower computational complexity. Overall, the implications of perfoming an adaptive DMRS/power ajustement within the 3GPP standard are significant in term of performance improvement. Specifically, the  {\em User Equipment} (UE) could vary  the power allocation between the  DMRS and data. This flexibility in  power allocation is transparent to the receiver.

\balance
\section{Conclusions}
In this paper, we presented a comprehensive analysis and the  performance enhancement of short block length channel detection incorporating  training information.
By investigating the receiver structure,  specifically the estimator-correlator, it was shown that the non-coherent term, which is often disregarded in conventional detection metrics, results in significant losses in terms of performance and sensitivity in typical operating regimes of 5G/6G systems.
Additionally, it was demonstrated that by employing an adaptive DMRS/data power adjustment, it is possible to reduce the performance loss gap which is amenable to a more sensitive and low complexity quasi-coherent receiver.
However, both of the  aforementioned ML detection strategies could result in substantial computational complexity when processing long bit length codes.  We proposed an innovative approach to tackle this challenge by introducing the principle of block/segment coding using First-Order RM Codes which  is amenable to low-cost decoding through block-based fast Hadamard transforms.
Thus, by incorporating an adaptive DMRS/data power adjustment technique, we were able to bridge/reduce the performance gap with respect to the conventional maximum likelihood receiver and attain high sensitivity, leading to a good trade-off between performance and complexity to efficiently handle small payloads.

\vspace{12pt}
\end{document}